\documentclass[twoside]{article}
\usepackage{amsmath, amsfonts, amssymb, amstext, wasysym}
\usepackage{graphicx}
\usepackage{aat}

\usepackage[paperheight=24cm,paperwidth=17cm,top=2cm,bottom=2cm,width=13cm,height=20cm]{geometry}

\usepackage[margin=0mm,labelfont=bf,labelsep=quad]{caption}

\begin{document}

\def\apj{Astrophys.~J. }
\def\aatr{Astron.~Astroph.~Trans. }
\def\aaps{Astron.~and Astrophys.~Suppl.~Ser. }
\def\pasp{Publ.~Astron.~Soc.~Pac. }
\def\gca{Geochim.~Cosmochim.~Acta. }
\def\aap{Astron.~Astrophys. }
\def\aspcs{ASP~Conf.~Ser. }
\def\asrep{Astron.~Rep. }
\def\nat{Nature. }
\def\apjl{Astrophys.~J.~Lett. }
\def\apjs{Astrophys.~J.~Suppl.~Ser. }
\def\aj{Astron.~J. }
\def\mnras{Mon.~Not.~R.~Astron.~Soc. }
\def\araa{Ann.~Rev.~Astron.~Astrophys. }
\def\jcp{J.~Chem.~Phys. }
\def\apss{Astrophys.~Space.~Sci. }
\def\prl{Phys.~Rev.~Lett. }
\def\phrva{Phys.~Rev.~A. }
\def\phlb{Phys.~Let.~B. }
\def\pf{Phys.~Fluids. }
\def\jgr{J.~Geophys.~Res. }
\def\cemda{Celest.~Mech.~Dyn.~Astr. }
\def\jcoph{J.~Comp.~Phys. }
\def\cophc{Comput.~Phys.~Commun. }
\def\phpl{Physics~of~Plasmas. }
\def\pasj{Publ.~Astron.~Soc.~Jpn. }
\def\jrasc{J.~R.~Astron.~Soc.~Can. }
\def\cemec{Celest.~Mech. }
\def\pasau{Proc.~Astron.~Soc.~Aust. }
\def\puasau{Publ.~Astron.~Soc.~Aust. }
\def\jasa{J.~Acoust.~Soc.~Am. }
\def\jfm{J.~Fluid~Mech. }
\def\cajph{Can.~J.~Phys. }
\def\mitag{Mitt.~Astron.~Ges. }
\def\bain{Bull.~Astron.~Inst.~Neth. }
\def\epsl{Earth~Planet.~Sci.~Lett. }
\def\ibvs{Inf.~Bull.~Variable~Stars. }
\def\arep{Astr.~Rep. }
\def\phr{Phys.~Rep. }
\def\astl{Astron.~Letters. }
\def\sci{Science. }
\def\jqsrt{J.~Quant.~Spectrosc.~Radiat.~Transfer. }
\def\emp{Earth,~Moon~and~Planets. }
\def\icar{Icarus. }
\def\pss{Planet.~Space~Sci. }
\def\qjras{Q.~J.~R.~Astron.~Soc. }
\def\nimpa{Nucl.~Instrum.~Methods~Phys.~Res.,~Sect.~A. }
\def\soph{Sol.~Phys. }
\def\lnm{Lect.~Notes~in~Math. }
\def\an{Astron.~Nach. }
\def\aph{Astroparticle~Physics. }
\def\adspr{Adv.~Space~Res. }
\def\geoj{Geophys.~J. }
\def\bamass{Bull.~Am.~Astron.~Soc. }
\def\rmxaa{Rev.~Mex.~Astron.~Astrofis. }
\def\aapr{Astron.~Astrophys.~Rev. }
\def\acp{Atmosphere~Chem.~Phys. }
\def\ssrv{Space~Sci.~Rev. }
\def\jmph{J.~Math.~Phys. }
\def\rvmps{Rev.~Mod.~Phys.~Suppl. }
\def\rvmp{Rev.~Mod.~Phys. }
\def\prd{Phys.~Rev.~D. }
\def\nuphs{Nuc.~Phys.~B~Proc.~Suppl. }
\def\nuphb{Nuc.~Phys.~B. }
\def\skytel{Sky~Telesc. }

\thispagestyle{myfirst}
\setcounter{page}{50}
\mylabel{35}{62}
\mytitle{Empirical dependencies for irregular dwarf galaxies}
\myauthor{K. Tillaboev$^{1,*}$, I. Tadjibaev$^{1,2}$, N. Otojanova$^{2}$}
\myaddress{
$^1$ National University of Uzbekistan, 100174, Tashkent, Uzbekistan, 4 University str.\\
$^2$ Chirchik State Pedagogical University, Chirchik, 104 A.Temur str.
}
\mydate{\today}
\myabstract{In this work, we search for empirical relationships for irregular dwarf galaxies using a collection of observational data published in the NASA and HyperLeda databases. Using a complex statistical analysis, we calculated correlation coefficients between various physical parameters of irregular dwarf galaxies. We found strong correlations between mass and distance, apparent stellar magnitude and distance, mass and absolute magnitude, redshift, and mass for all types of irregular dwarf galaxies as well as for their individual types.}
\mykey{irregular dwarf galaxies, empirical dependencies, correlation, heliocentric velocity, redshift, HyperLeda database, catalog.}

\mbox{}\footnotetext{${^{*}}$Email: tillaboyev\_k@nuu.uz}

\section{Introduction}

In the universe, among the different types of galaxies, there are dwarf galaxies consisting of a few billion stars with luminosities less than $10^9 L_\odot$ (Kutlimuratov et al. 2020). Most often dwarf galaxies are those whose absolute stellar magnitude does not exceed $-16^m$. Therefore, it is very difficult to observe dwarf galaxies at large distances (Zasov, 1984).

Currently, many papers have been published on dwarf galaxies. For example, besides, the first catalog of dwarf galaxies was published in the paper (van den Bergh, 1959). Van den Bergh's catalog (van den Bergh, 1959) includes all low-surface brightness galaxies that are present on the Palomar Sky Survey. In the catalogue (Mateo, 1998), the author discusses detailed properties of dwarf galaxies, including (a) the integral photometric parameters and optical structures of these galaxies, (b) the content, nature, and distribution of their interstellar medium, (c) their abundances of heavy elements, (d) their complex and diverse star formation histories, (e) their internal kinematics, emphasising the relevance of these galaxies to the `dark matter problem' and alternative interpretations, and (f) evidence of past, current and future interactions of these dwarfs with other galaxies in the Local Group and beyond. This paper (McConnachie, 2012) presents a comprehensive analysis of the positional, structural, and dynamical parameters of over 100 dwarf galaxies located within 3 Mpc of the Sun. These galaxies cover a wide range of environments, from satellites of the Milky Way and Andromeda to isolated systems in the neighborhood of the Local Group. The study presents homogeneous data on key observational parameters - distances, velocities, magnitudes, metallicities and structural features - summarised in continuously updated tables. The paper explores the spatial structure and subgroup membership in the Local Group, examining the morphological diversity and orbital history of these galaxies. In addition, the scale relations and trends in stellar metallicity are investigated, including the existence of a possible lower limit to the central surface brightness and metallicity at faint magnitudes.

In (Poulain et al, 2021), the sample consists of 2210 dwarfs. The authors define the nucleus as a compact source close to the photocentre of the galaxy (within $0.5R_e$), which is also the brightest such source within the effective radius of the galaxy. We performed a morphological analysis by modeling the two-dimensional surface brightness profile in g-band images of both galaxies and nuclei. The study (Guo et al, 2024) presents a comprehensive catalogue of 3459 nearby blue star-forming dwarf galaxies derived from SDSS DR16, focusing on the evolution of their metallicity, an area less studied in dwarf galaxy studies. The authors apply strict selection criteria based on redshift, stellar mass, absolute magnitude, optical diameter, and emission line signal-to-noise ratio to create a sample from the higher-mass end of the dwarf galaxy spectrum. Despite the range of metallicities, the average metallicity was found to be about 0.2 dex below solar, indicating relatively mature chemical evolution.

But the information in such catalogues is not sufficient for a complete physical study of dwarf irregular galaxies. Especially there is no catalogue on separate morphological types of dwarf galaxies. Based on the study of existing catalogues, we have created a special summary catalogue of dwarf galaxies containing 413 objects. Based on this catalogue, we calculated correlation coefficients between the main physical characteristics of irregular dwarf galaxies.

\section{Find empirical dependences}

  Using data from the Special summary catalogue of dwarf galaxies containing 413 objects, we calculated correlation coefficients between the main physical characteristics of irregular dwarf galaxies.

\begin{figure}[h]
\centering
\includegraphics[width=1\textwidth]{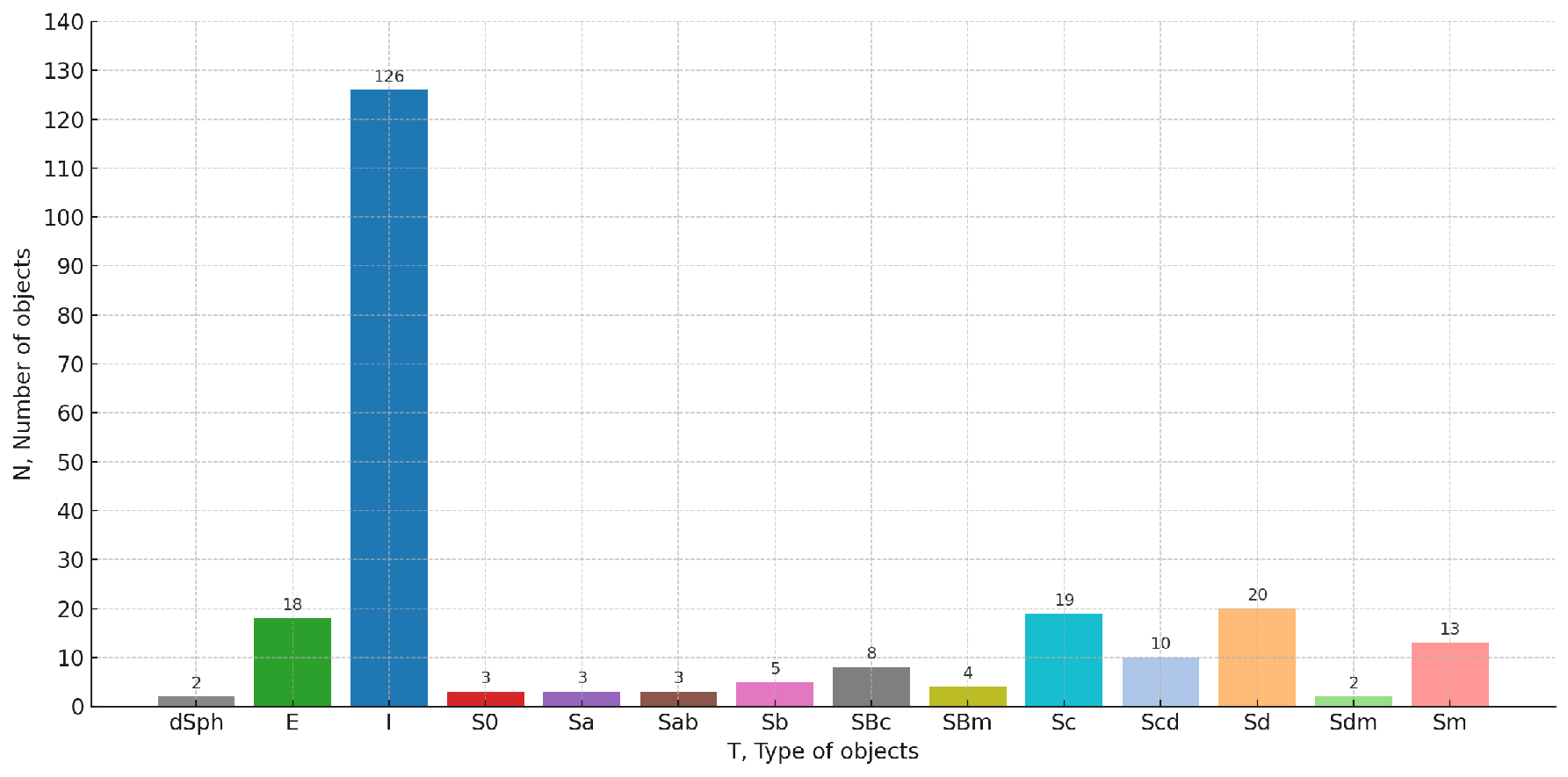}
\caption{Histogram of the distribution of dwarf galaxies by morphological types.}
\label{fig:histogram}
\end{figure}

According to the data in Figure 1, the majority of the objects are irregular morphological type dwarf galaxies. The number of spiral dwarf galaxies is also high, and the histogram shows several of their types.

Based on our data, we found a linear dependence of some quantities for irregular dwarf galaxies. When the relationship between mass and distance (correlation 0.95) was studied, the following relationship was obtained:
\begin{equation}
\log(M/M_{Sun}) = 0.0230(\pm0.0006)\cdot D + 0.7840(\pm0.0504)    
\end{equation}
It is evident that with increasing distances the mass of irregular dwarf galaxies increases.

An empirical relationship between apparent stellar magnitude and distance was also found (the correlation coefficient is 0.58):
\begin{equation}
m_B= 0.0206(\pm0.0025)\cdot D + 16.0010(\pm0.1953)
\end{equation}

The dependence (2) shows that the brighter or larger the stellar magnitude of the galaxy, the further away they are. The errors in determining the coefficients are quite acceptable.

Further, we have analyzed the statistical dependence of the galaxy mass on the absolute stellar magnitude of the galaxy itself. We have found the following empirical dependence by the least squares method (correlation equal to -0.58):
\begin{equation}
\log(M/M_{Sun}) = -0.6670(\pm0.0784)\cdot M_v -1.4975(\pm1.2965)
\end{equation}
As can be seen, the logarithmic mass of galaxies increases with increasing absolute magnitude within the linear law.

Obviously, the values of the absolute stellar magnitude should correlate with the redshift of the mass. Calculation of this empirical dependence shows that (correlation 0.95):
\begin{equation}
z = 0.0094(\pm0.0002) \cdot \log(M/M_{Sun}) -0.7270(\pm0.0025) 
\end{equation}
It can be seen that here the coefficient determination errors are quite acceptable.

The above results were calculated for all irregular dwarf galaxies in the list. In order to compare these results, the irregular dwarf laws are in the list of dividends into your parts. The first part included only those with type codes equal to magnitude 10, and the second part included those with type codes between magnitudes 9.5 and 9.9. The results below are for galaxies with type codes equal to 10.

Relationship between mass and distance (correlation 0.98):
\begin{equation}
\log(M/M_{Sun}) =0.0250(\pm0.0004) \cdot D + 7.6860(\pm0.1490)
\end{equation}

Relationship between apparent magnitude and distance (correlation 0.62):
\begin{equation}
m_B= 0.0140(\pm0.0004) \cdot D+16.575(\pm0.148)
\end{equation}

Relationship between mass and absolute magnitude   (correlation -0.63):
\begin{equation}
\log(M/M_{Sun}) = -0.6670(\pm0.0784) \cdot M_v -1.4975(\pm1.2965)
\end{equation}

Relationship between redshift and mass  (correlation 0.98):
\begin{equation}
z = 0.00093(\pm0.0001) \cdot \log(M/M_{Sun}) -0.0716(\pm0.0016)
\end{equation}

The results below are for galaxies with type codes between magnitude 9.5 and 9.9.

Relationship between mass and distance (correlation 0.82):
\begin{equation}
\log(M/M_{Sun}) = 0.0504(\pm0.0088) \cdot D + 7.5715(\pm0.2227)
\end{equation}

Relationship between  mass and absolute magnitude (correlation -0.91):
\begin{equation}
\log(M/M_{Sun}) = -0.3837(\pm0.0422) \cdot M_v -2.2864(\pm0.7116)
\end{equation}

Relationship between redshift and mass   (correlation 0.81):
\begin{equation}
z = 0.0031(\pm0.0005)\cdot \log(M/M_{Sun}) -0.0223(\pm0.0051)
\end{equation}

\textup {\itshape Mass vs. Distance.\/}The correlation increases slightly in the type code = 10 group (0.95 $\to$ 0.98), indicating a stronger relationship between mass and distance in this subset. Conversely, the correlation decreases significantly in the type code 9.5–9.9 group (0.95 $\to$ 0.82), suggesting a weaker relationship.

\textup{\itshape Apparent  magnitude vs. Distance.\/} The correlation improves slightly in the type code = 10 group (0.58 $\to$ 0.62). For the type code 9.5–9.9 group, the correlation is not observed.

\textup{\itshape Mass vs. Absolute magnitude.\/} The negative correlation strengthens slightly in the type code = 10 group ($|-0.58| \to |-0.63|$). In the type code 9.5–9.9 group, the correlation increases dramatically ($|-0.58| \to |-0.91|$), indicating a significantly stronger negative relationship between visual magnitude and mass.

\textup{\itshape Redshift vs. Mass.\/} The correlation increases in the type code = 10 group (0.95 $\to$ 0.98), reflecting a stronger relationship. In contrast, it decreases in the type code 9.5–9.9 group (0.95 $\to$ 0.81), indicating a weaker connection.

\section{Conclusions}

This study provides a detailed examination of the empirical dependencies among physical parameters of irregular dwarf galaxies, revealing significant correlations that enhance our understanding of their properties and behavior. The strong correlations between mass and distance (up to 0.98 in specific subsets) and mass and redshift (up to 0.98) suggest that these galaxies follow distinct structural and kinematic patterns, particularly in morphologically homogeneous groups (type code = 10). These findings indicate that morphological classification plays a critical role in shaping the physical relationships within irregular dwarf galaxies. Our results contribute to the broader framework of galaxy evolution by providing empirical constraints on the properties of low-luminosity systems. Future studies could extend this work by incorporating multi-wavelength observations or dynamical modeling to further elucidate the formation and evolution of irregular dwarf galaxies. These efforts will be crucial for refining our understanding of their role in the hierarchical structure of the Universe.

\section{References} 

\noindent
Guo, Y., Sengupta, C., Scott, T. C., Lagos, P., \& Luo, Y. 2024, MNRAS, 528, 6593

\vspace{2mm}
\noindent
Kutlimuratov, S., Nuritdinov, S., \& Tadjibaev, I. 2020, Uzbek Journal of Physics, 22, 203

\vspace{2mm}
\noindent
Makarov, D., Prugniel, P., Terekhova, N., Courtois, H., \& Vauglin, I. 2014, A\&A, 570, A13

\vspace{2mm}
\noindent
Mateo, M. L. 1998, ARA\&A, 36, 435

\vspace{2mm}
\noindent
McConnachie, A. W. 2012, AJ, 144, 4

\vspace{2mm}
\noindent
Poulain, M., Marleau, F. R., Habas, R., Duc, P.-A., Sánchez-Janssen, R., Durrell, P. R., \& Fensch, J. 2021, MNRAS, 506, 5494

\vspace{2mm}
\noindent
Tadjibaev, I., Tillaboev, K., \& Otojanova, N. 2023, EUREKA: Physics and Engineering, 2023(6), 3–11.

\vspace{2mm}
\noindent
van den Bergh, S. 1959, Publ. David Dunlap Obs., Univ. Toronto, 2, 5, 147

\vspace{2mm}
\noindent
Zasov, A. V. 1984, \textit{Dwarf Galaxies}, Znanie, Moscow

\end{document}